\documentclass[10pt,a4paper]{article}
\usepackage[explicit]{titlesec}
\usepackage{hyphenat}
\usepackage{ragged2e}
\usepackage{booktabs}
\usepackage{chemscheme}
\usepackage{setspace}
\usepackage{bm} % for the \boldsymbol
\usepackage{textcomp}
\RaggedRight
\justifying
\usepackage{float}
\floatstyle{plaintop}
\restylefloat{table}
\usepackage[T1]{fontenc}
\usepackage{amsmath, amsthm}
\usepackage{graphicx}
\usepackage[utf8]{inputenc}
\usepackage{epsfig}
\usepackage{subcaption}
\usepackage{soul}
\setul{.6pt}{.4pt}
\usepackage[version=4]{mhchem}
\usepackage{geometry}
\usepackage{hyperref}
\usepackage{cleveref}
\usepackage{multicol}
\usepackage{fancyhdr}

\makeatletter
\renewcommand\tagform@[1]{\maketag@@@ {\ignorespaces {\footnotesize{\textbf{Equation}}} #1.\unskip \@@italiccorr }}
\makeatother
\titleformat{\section}
  {\normalfont}{\thesection}{1em}{\MakeUppercase{\textbf{#1}}}
\titlespacing\section{0pt}{2ex}{0ex}
\titleformat{\subsection}
  {\normalfont}{\thesubsection}{1em}{\textbf{#1}}
  \titlespacing{\subsection}{0pt}{1.5ex}{0ex}
\titleformat{\subsubsection}
  {\normalfont}{\thesubsubsection}{1em}{\textit{#1}}
  \titlespacing{\subsubsection}{0pt}{1ex}{0ex}

\makeatletter
\newcommand\sixteen{\@setfontsize\sixteen{16pt}{6}}
\renewcommand{\maketitle}{\bgroup\setlength{\parindent}{0pt}
\begin{flushleft}
\vspace{-.375in}
\sixteen\bfseries \@title
\medskip
\end{flushleft}
\textit{\@author}
\egroup}
\makeatother
\usepackage[sort&compress]{natbib}
\makeatletter
\renewcommand\@biblabel[1]{\textbf{#1.}\hfill}
\makeatother

\bibpunct{}{}{,~}{s}{,}{,}
\title{The Role of a Diluent in Deformation-Induced Bonding of Glassy Polymer Bidisperse Blends}
\author{\begin{center}Ajay Vallabh\textsuperscript{1*} and John Tsavalas\textsuperscript{2*}\end{center}
\textsuperscript{1}Department of Mechanical Engineering, University of New Hampshire, Durham, NH 03824 USA \newline
\textsuperscript{2}Department of Chemistry, University of New Hampshire, Durham, NH 03824 USA \vspace{1em}
\begin{center}{\textsuperscript{*}Corresponding authors: ajay.vallabh@unh.edu, john.tsavalas@unh.edu}\end{center}
}
\date{June 2024}

\begin{document}
%\begin{doublespace}
\maketitle
\begin{abstract}
\begin{center} 
Bonding between polymers below the glass transition temperature through molecular-scale dilatation (or densification)-based interdiffusion of macromolecules has recently been introduced. In this mechanism, short timeframe plastic deformation enables polymer chains to interdiffuse and form entanglements at the interface, facilitating rapid bonding below the glass transition temperature ($T_g$). Here, we are addressing the role of a lower molecular weight diluent in bonding polymer interfaces of bidisperse blends through deformation-induced bonding (DIB) at temperatures well below both the surface and bulk glass transition temperatures, $T_g^s$ and $T_g^b$, respectively, by using molecular simulations. These simulations reveal that addition of the diluent ($\phi\le$20\%) drastically enhances the number of chain-ends at the interfacial region compared to a pure glass sample ($\phi=0\%$) during deformation below $T_g^s$, which improves the possibility of opposite side entanglement formation. The changes in stress-strain response of debonded samples correlate with the normalized entanglement density. Likewise, the maximum interfacial fracture energy $G_{I,max}$ of debonded samples is correlated with the diluent concentration ($\phi$), below $T_g^s$. Furthermore, the optimization of material and process conditions for DIB has yielded a notable advancement for the conditions tested here: achieving a higher bonding strength, approximately one-third of the bulk, all while remaining below $T_g$. 
\end{center}
\end{abstract}
%Simulations reveal that adding plasticizer (0\% to 40\%) enhances the molecular mobility of monomer beads up to twofold in the glassy state and up to two orders of magnitude during deformation compared to a pure glass sample.
%number of host polymer chain-end enhancements at the interface during deformation, which is contingent on

\section{Introduction}

Plasticizer is a low-molecular-weight additive commonly added to polymeric materials to change their thermophysical and mechanical properties, such as glass transition temperature ($T_g$), transparency, material stiffness, ductility, and toughness in the glassy state. These characteristics make them suitable for a wide range of applications. Oftentimes these additives are leveraged to blend with glassy polymers to reduce the composite glass transition so as to cast stable polymer films. These days, sequential film formation is growing in popularity where multilayer films are appropriate for applications such as multi-layer drug tablets \cite{Trout2015, padhye2023mechanics, padhye2015sub}, packaging \cite{zhang2006interfacial}, medical patches for wound healing \cite{naseri2022review},  and fabrication of smart sensor patches \cite{verdel2023use} for monitoring body data, such as blood glucose level or heartbeat of patients, to name a few; each of these may also include the use of different bonding processes. In most cases, plasticizers are applied so as to reduce the $T_g$ to be below that of the filming temperature.

The most common polymer film bonding processes include calendering\cite{nutter1991calender}, ultrasonic welding\cite{troughton2008handbook}, extrusion coating \cite{kang2009extrusion}, hot wedge welding\cite{jana2011assembling}, and hot air welding\cite{jana2011assembling}, etc., which work on the principle of conventional heat-based thermal fusion, where two polymer substrates are heated above their glass transition temperature $T_g$ and subsequently pressed together under moderate pressure. This process enables the interdiffusion of polymer chain segments across the interface, leading to the formation of entanglements and the creation of a robust bonded interface within an experimentally feasible timeframe\cite{jud1981fracture,voyutskii1963role,kausch1989polymer,prager1981healing,jud1979load,de1981formation,wool1981theory,wool1995polymer}. Experiments\cite{kline1988polymer} confirm that in the thermal fusion bonding process, the interfacial fracture toughness $G_c$ and shear strength $\sigma_s$ of the polymer interface both grow monotonically\cite{kline1988polymer}, follow the power law $t^{1/2}$ and $t^{1/4}$, respectively, and bonding takes place only above $T_g$.

In our recent experiments\cite{padhye2021deformation}, we prepared various blends of hydroxypropyl methylcellulose (HPMC) and polyvinyl alcohol (PVA) with polyethylene glycol (PEG-400) as a low molecular weight additive to cast stable films and bond them \textit{below} their $T_g$ by using a new deformation-induced bonding (DIB) phenomena\cite{padhye2017new}. Here, the PEG-400 did indeed plasticize each blend's composite $T_g$, however not to a value lower than the bonding temperature. In this way, the PEG-400 was leveraged as a minor-fraction lower molecular weight diluent to the respective major-fraction chains. In DIB, two polymer substrates are brought into contact and deformed through bulk compression. Plastic deformation triggers and enhances chain segments' mobility \cite{capaldi2002enhanced,capaldi2004molecular}, allowing them to diffuse and form entanglements across the interface to form a strongly bonded interface in a only a few seconds\cite{padhye2017new} at ambient temperature (and well below the system composite $T_g$). Experiments\cite{padhye2017new,padhye2021deformation} showed film samples bonded by bulk plastic compression exhibit non-monotonic bonding strength, which is quantified as the interfacial fracture toughness $G_c$ of different samples corresponding to various levels of plastic strain.  

To complement that work, we have used molecular dynamics (MD) simulations\cite{padhye2022dilatational} to reveal the mechanism of DIB, and confirmed that the enhanced molecular scale dilatation (or densification) combined with faster molecular mobility during plastic deformation allows for polymer chain segment interdiffusion even at temperatures below $T_g$. This phenomenon leads to the formation of kinks\cite{argon1973theory}, entanglements and subsequent disentanglements, resulting in non-monotonic bonding strength. Furthermore, simulation results highlighted the internal structure of the glass, confirming the absence of liquidity (or free volume) enhancement during deformation. Instead, we observed a redistribution of pre-existing liquidity from one region to another, leading to the formation of density-rich regions (densification) and low-density regions (dilatation).

However, the initial MD model\cite{padhye2022dilatational} was built for an amorphous homopolymer, excluding consideration of the effects resulting from the addition of a diluent under plane strain compression conditions. Here, the diluent was used to cast highly ductile polymer films by tuning the density of entanglements in experiments\cite{padhye2021deformation}. Recently, Zhang et al.\cite{zhang2021thickness} used MD simulations to examine the mechanical properties of chemically identical glassy polymer bidisperse blends. They concluded that by changing the molecular weight of the major and minor volume fractions of bidisperse blends, the film thickness and desired mechanical failure, toughness, and stress-strain response could be achieved. 
 
The current study uses MD simulations to elucidate the role of a diluent in bonding between two interfaces and the impact of that on bonding strength. Here, we considered chemically identical host polymer (major fraction) and diluent (minor fraction) to make glassy polymer films of bidisperse blends as a function of diluent volume fraction and molecular weight. Specifically, two types of bidisperse systems were prepared: (i) a configuration comprising a highly entangled host polymer paired with a low molecular weight \textit{unentangled} diluent polymer (with kink $Z\le1$), and (ii) another configuration comprising a moderately entangled host polymer paired with a medium molecular weight \textit{weakly entangled} diluent (with kink $Z>1$). Subsequently, we varied the diluent concentration ($\phi$ by volume) for each case and conducted entanglement analyses, including surface ($T_g^s$) and bulk ($T_g^b$) glass transition temperature measurements, radial distribution function calculations, uniaxial tensile tests, and toughness measurements. These investigations shed light on how diluents (the entanglement network of different blends thereof with host polymer) influence the thermophysical and mechanical properties of glassy polymer films.

Plane strain compression was performed to bond the glassy polymer films of bidisperse blends at temperatures below $T_g$; the bonding strength of the resulting samples was subsequently measured by conducting a uni-axial tensile test and quantified as interfacial fracture energy $G_I$. The roles of molecular mobility, dilatation (or densification)-based interdiffusion, and number of chain-ends were analyzed to correlate the variation in bonding strength of different bonded bidisperse blend samples during and after deformation.       
 
\section{THEORETICAL BASIS}

To model bidisperse glassy polymer films under plane strain compression, we began with similar models to our former work \cite{padhye2022dilatational}, where the comprehensive details of bonding under bulk compression and debonding through uniaxial tensile testing can be found. All the simulations employed a coarse-grain model that can capture the properties of bidisperse blends\cite{grest2022entropic} of diluent (D) and host polymer (H). Samples were prepared by following the Kramer-Grest bead-spring model\cite{kremer1990dynamics}. All the beads interact with each other through a truncated and shifted Lennard-Jones (LJ) potential,
\begin{equation}
U_{LJ}(r)=4u_o\left[\left(\frac{a}{r}\right)^{12}-\left(\frac{a}{r}\right)^{6}\right]    
\label{Eq1}
\end{equation}

\begin{equation}
U_{LJTS}(r)= \begin{cases}
    U_{LJ}(r)-U_{LJ}(r_c) \;\;\;\;\;\;r\le r_c \\
    0  \;\;\;\;\;\;\;\;\;\;\;\;\;\;\;\;\;\;\;\;\;\;\;\;\;\;\;\;\;\;\;r> r_c
\end{cases}
\end{equation}

where $u_o$ and $a$ indicate energy and the size of each bead (i.e. each monomer unit), respectively, and $r$ represents the distance between two particles. Initially, $r_c$, i.e., the cutoff radius, was set to be $2^{1/6}a$ to remove overlapping of chains for initial polymer blend configuration; for later simulations, $r_c$ was set to be $2.5a$ to mimic a full LJ interaction (comprising both repulsion and attraction).   

We set equal bead diameter $\sigma_D$=$\sigma_H$=$a$, equal binding energy $\epsilon_D$=$\epsilon_H$=$u_o$, and equal mass $m_D$=$m_H$=$m$ for diluent and host polymer chains. Here, equal values of epsilon helps to prevent phase separation during quenching as the Flory parameter\cite{grest2022entropic} $\chi$=$z\frac{\Delta\epsilon}{KT}$ approaches zero, where $\Delta\epsilon=\epsilon_{DH}-\frac{\epsilon_D}{2}-\frac{\epsilon_H}{2}$. Cross interactions $\epsilon_{DH}$ and $\sigma_{DH}$ between polymer beads and diluent beads were determined using the Lorentz–Berthelot combination rules \cite{lorentz1881ueber,berthelot1898melange}, $\epsilon_{DH}=\sqrt{\epsilon_{D}\times\epsilon_{H}}$ = $u_o$ and $\sigma_{DH}=\frac{\sigma_{D}+\sigma_{H}}{2}$ = $a$.

Intraparticle interactions between connected beads of a chain are defined by the standard unbreakable finitely extensible non-linear elastic (FENE) potential\cite{kremer1990dynamics}.
\begin{equation}
U_{FENE}(r)=-\frac{1}{2}kR_o^2ln\left[1-\left(\frac{r}{R_o}\right)^{2}\right]   
\end{equation}
where standard values of the spring constant $k=30u_o/a^{2}$, and the maximum bond extension parameter $R_o=1.5a$ are used. Later, the FENE potential was replaced with a quartic potential\cite{ge2013molecular} that allowed for bond breaking as the tensile tests proceeded. To obtain the same equilibrium bond length as $U_{FENE}$, the following relationship was constructed 
\begin{equation}
U_Q(r)=K(r-R_c)^{2}(r-R_c)(r-R_c-B)+U_o
\end{equation}
where $K=2351 u_o/k_B$, $B=-0.7425a$, $R_c=1.5a$, and $U_o=92.74467u_o$.

All the simulations were performed on LAMMPS software\cite{thompson2022lammps}, where simulation quantities were expressed in the lj unit of LAMMPS with the aid of molecular diameter $a$, binding energy $u_o$, and the characteristic time $\tau=a\left(m/u_o\right)^{\frac{1}{2}}$. Direct mapping of our model to a specific polymer is not in the scope of this work. Nevertheless, we can offer approximate mappings, such as $u_o/a^3\sim$ 50 MPa and $u_o/a^2\sim$ 25 $mJ/m^2$ for stress and fracture energy units, respectively, drawing from prior simulations \cite{ge2013molecular, ge2014healing}. The velocity Verlet algorithm was used to solve equations of motion with a time step of $\delta t$ = $0.0005\;\tau$ to $0.01\;\tau$. 

Initial polymer blend samples were categorized based on molecular weight ($M_w$) and entanglement length ($N_e$)\cite{everaers2004rheology} into two distinct cases. In Case 1, the blend comprised a highly entangled host polymer accompanied by unentangled polymer diluent, consisting of a total of 250,000 monomer beads with a mass denoted by '$m$'. Meanwhile, Case 2 involved moderately entangled host polymer chains with weakly entangled diluent chains, composed of 163,840 monomer beads. Table \ref{tab:ChainLength} illustrates the geometrical features of the polymer blend samples for each respective category.
 
 \begin{table}[H]
    \caption{\textbf{Geometrical features of different polymer samples}}
    \centering
    \begin{tabular}{|c|c|c|c|c|c|c|}
    \hline
       Sample & Diluent& Number of  & Host polymer  & Number of & Diluent& Entanglement\\
       No. & concentration & host polymer & chain length & diluent& chain  &length \\ 
       & & chains& & chains & length &  \\
       &($\phi$ by volume) & $M_h$ & $N_h$ & $M_d$& $N_d$& $N_e$ \\ \hline
        \multicolumn{7}{|l|}{Case 1: Highly entangled host polymer with unentangled diluent polymer} \\\cline{1-7}
        1 & 0\% & 500 & 500 & 0 & 8 & 82\\ \hline
        2 & 10\% & 450 & 500 & 3125& 8&-\\ \hline
        3 & 20\% & 400 & 500 & 6250 & 8&-\\ \hline
        4 & 40\%&300 &500 &12500&8&-\\ \hline
        \multicolumn{7}{|l|}{Case 2: Moderately entangled host polymer with weakly entangled diluent polymer} \\\cline{1-7}
        5 & 0\% & 1280 & 128 & 0 & 64& 56\\ \hline
        6 & 10\% & 1152 & 128 & 256 & 64&-\\ \hline
        7 & 20\% & 1024 & 128 & 512 & 64&-\\ \hline
        8 & 40\%& 768 & 128 &1024& 64&-\\ \hline
    \end{tabular}
    \label{tab:ChainLength}
\end{table}

The melt state\cite{auhl2003equilibration} of all these samples was prepared by running a long-run simulation ($1$x$10^7\tau$) at a constant density of $\rho$ = 0.85$a^{-3}$ and temperature ($T$ = 1.0 $u_o/k_B$) with periodic boundary conditions in all directions. Then, fixed boundary conditions were imposed in the z-direction, and samples were equilibrated at $T$ = 1.0 $u_o/k_B$, under the NPT ensemble. The samplepressure was kept constant at $P$ = 0 with a damping coefficient of $P_{damp}$ = $1000\times\delta t$, by permitting expansion or contraction along the x-direction. The equilibrated samples of Case 1 were quenched from $T$ = 1.0 $u_o/k_B$ to $T$ = 0.3 $u_o/k_B$, which is below the glass transition temperature $T_g$ $\approx$ 0.45 $u_o/k_B$, at a quenching rate of $\Dot{T}$ = 2$\times10^{-3}u_o/k_B\tau$ to achieve a glassy state. Quenching was performed under an NPT ensemble, where sample pressure $P$ = 0 was maintained by allowing expansion and contraction in the x and y directions, with damping coefficient $P_{damp}$ = $1000\times\delta t$. After that, the samples were again equilibrated for $0.1\;M\tau$ at $T$ = $0.3\:u_o/k_B$, under NPT ensemble to remove residual stresses; here, $M$ abbreviates 1 million.

\begin{figure}[H]
\centering
    \includegraphics[scale=1]{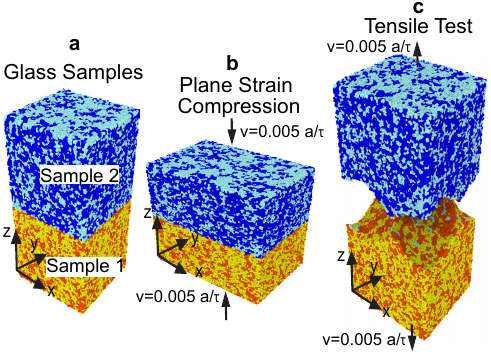}
    \caption{\textbf{Simulation setup of deformation-induced bonding and debonding.} \textbf{a}, polymer blend samples in the glassy state. \textbf{b}, plane strain compression test. \textbf{c}, uniaxial tensile test of the bonded interface. Here, dark blue and yellow colors represent host polymer chains, while sky blue and red colors represent diluent molecules.}
    \label{fig:SimulationSetup}
\end{figure}

Another set of glassy samples with 0\%, 10\%, 20\%, and 40\% diluent concentrations was generated by following the preceding steps. Then, two samples of the same concentration were placed one over the other without overlap at z = 0 in a simulation box to perform deformation-induced bonding between them (See Figure \ref{fig:SimulationSetup}a). After that, the system of two glassy samples of the same concentrations was compressed by moving the top and bottom fixed boundary walls with ramp velocity $v=0.005\;a\tau^{-1}$ toward z=0 along the z-axis, allowing the samples to extend in the x-direction while the y-direction remained constant to achieve plane strain compression (See Figure \ref{fig:SimulationSetup}b). The system was kept at $T$ = $0.3\:u_o/k_B$, and $P$ = 0 under an NPT ensemble, with $P_{damp}$ = $1000\times\delta t$ maintained in $P_{xx}$ during bonding. Six different bonded samples of each concentration were generated, corresponding to $5\%$-$\epsilon^p$, $10\%$-$\epsilon^p$, $15\%$-$\epsilon^p$, $20\%$-$\epsilon^p$, $25\%$-$\epsilon^p$, and $30\%$-$\epsilon^p$ plastic strain (or deformation time duration up to 4000 $\tau$). Then, each of these bonded samples was equilibrated for 0.1 $M\tau$ to ensure elastic recovery.

To measure the bonding strength of bonded samples, uniaxial tensile tests were performed in a region of height $Lz_o$ = 50 units around the center of the bonded interface (See Figure \ref{fig:SimulationSetup}c). Outside of this core region, the remaining material portions were rigidly held and moved in opposite directions at a constant velocity $v$ = $0.005\:a\tau^{-1}$. Nominal strain is defined as $\epsilon^p$ = $Lz/Lz_o-1$, where $Lz$ is the growing length between rigid portions. The temperature of the core region was maintained at $T$ = $0.3\:u_o/k_B$ under the NVT ensemble, with $T_{damp}$ = $100\times\delta t$ . 

For Case 2 samples, we carried out bonding experiments (plane strain compression test) at $T$ = 0.4 $u_o/k_B$, which is still below $T_g^b$ and near $T_g^s$. The bonding strength of these samples was measured by performing a uniaxial tensile test, as defined above, for a different characteristic chain length of diluent and thus the number of chains. Similarly, to obtain the bulk tensile response of each sample, identical procedures were then followed to perform the uniaxial tensile test at $T$ = 0.3 $u_o/k_B$ as for the bonded interfaces.

\section{RESULTS AND DISCUSSION}

\subsection{Entanglement analysis}

The Z1+ algorithm\cite{kroger2023z1+} was employed to determine the average entanglement per chain, denoted as <$Z$>, in the bidisperse polymer blends. This algorithm works on geometrical\cite{karayiannis2009combined} configurations and calculates kinks, defined as the point where polymer chains bend around neighboring chains, by reducing the chain configuration iteratively.

\begin{figure}[H]
\centering
    \includegraphics[scale=1.0]{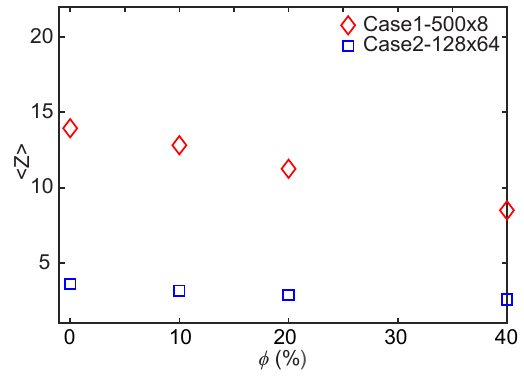}
    \caption{\textbf{Entanglement analysis.} Average entanglement per chain <$Z$> of bidisperse blend samples of Case 1 and Case 2 as a function of diluent concentration ($\phi$).}
    \label{fig:Z1entanglement}
\end{figure}

In our entanglement analysis, we excluded the lower-molecular-weight diluent chains and chains with kink $Z\approx1$ from the sample configuration. This decision was made because very low-molecular-weight diluent chains (e.g. $N_d$ = 8; i.e. Case 1) or chains with only one kink (possible in both Case 1 or 2) are incapable of contributing to mechanical loading\cite{bukowski2021load}. Only chains that meet the criteria $N\geq2N_e$ can contribute effectively to stable craze formation and mechanical loading\cite{rottler2003growth,baljon2001simulations}. Figure \ref{fig:Z1entanglement} shows that average entanglement per chain <$Z$> decreases as the diluent concentration ($\phi$) increases for both Case 1 and Case 2 samples, which is consistent with previous studies for bidisperse polymer blends\cite{zhang2021thickness}.   

\subsection{Glass transition temperature and Radial distribution function} 

\begin{figure}[H]
\centering
    \includegraphics[scale=0.9]{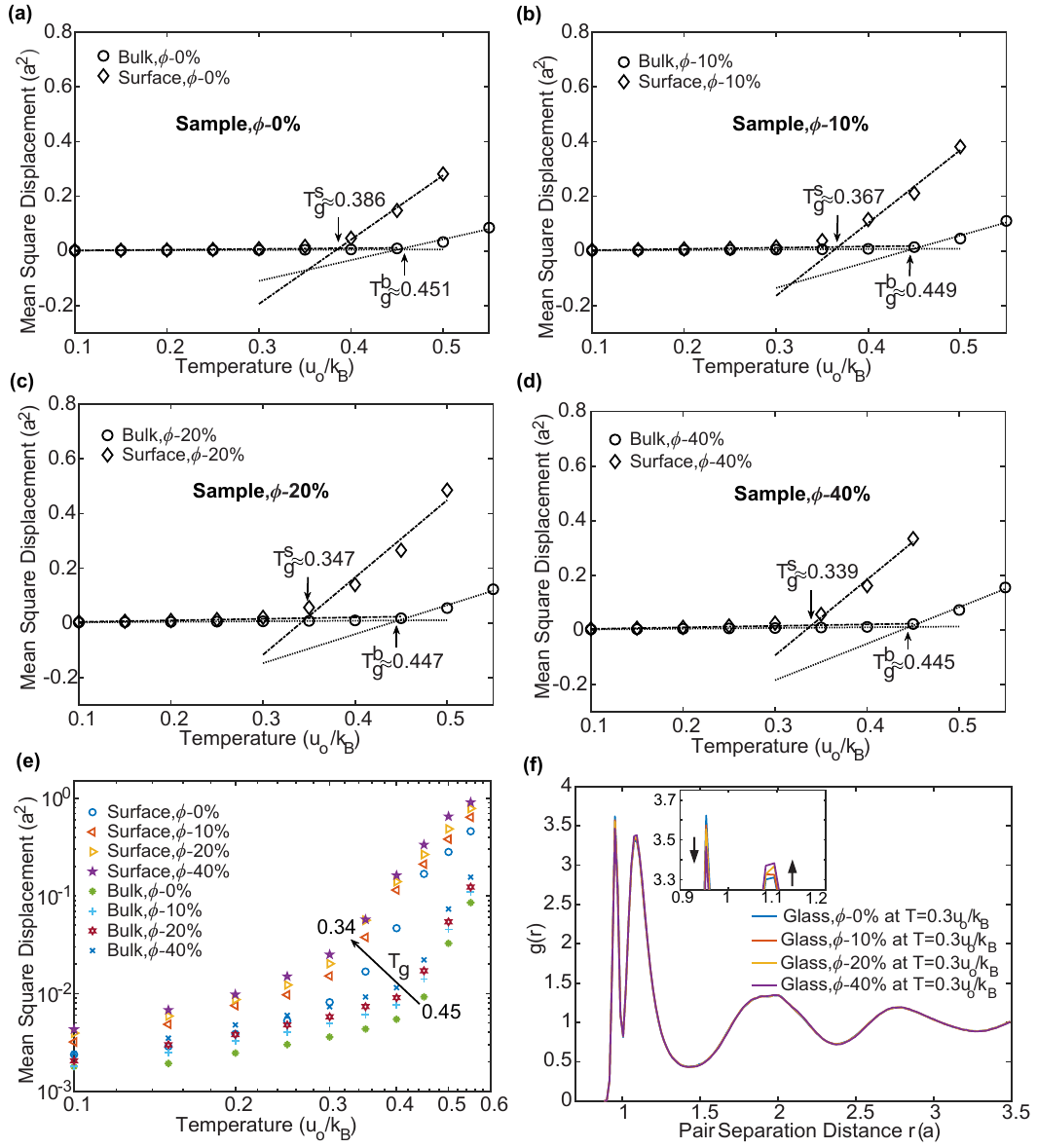}
    \caption{\textbf{Glass transition temperature and radial distribution function plots for bidisperse polymer blend samples of Case 1.} \textbf{a-e}, Estimation of the surface and bulk glass transition temperatures, denoted as $T_g^s$ and $T_g^b$ respectively, for $\phi$ = 0\%, 10\%, 20\%, and 40\% diluent containing samples by analyzing the mean square displacement (MSD) as a function of temperature. \textbf{f}, Plot of RDF comparing ($\phi$ = 0\%) pure glass sample with ($\phi$ = 10\%, 20\%, and 40\%) diluent containing bidisperse glass samples.}
    \label{fig:GlassTransition}
\end{figure}

Due to the presence of low-molecular-weight diluent chains in the bidisperse films, the segmental dynamics of the film are affected as the diluent chain concentration $\phi$ (\%) increases. To better understand the diluent effect on segmental mobility in the glassy state, the free surface ($T_g^s$) and interior bulk ($T_g^b$) glass transition temperatures were calculated by generating mean-square-displacement (MSD) vs. temperature plots using a standard procedure\cite{morita2006study,padhye2022dilatational}. The glass transition temperature of the free surface ($T_g^s$) and interior bulk ($T_g^b$) were estimated to be $T_g^s$= $0.386$, $0.367$, $0.347$, $0.339\:u_o/k_B$ and $T_g^b$ = $0.451$, $0.449$, $0.447$, $0.445\:u_o/k_B$, respectively, at diluent concentrations of $\phi$ = 0\%, 10\%, 20\% and 40\%, as shown in Figures \ref{fig:GlassTransition}a-e. 

Both $T_g^s$ and $T_g^b$ exhibit a decrease in value relative to $\phi$, which can be elucidated through a radial distribution function (RDF) plot, employed to capture the internal molecular structure of samples. In Figure \ref{fig:GlassTransition}f, the corresponding RDF plot illustrates $\phi$ = 0\% for the pure glass and $\phi$ = 10\%, 20\%, and 40\% plasticized (but still glassy) samples at $T$ = 0.3 $u_o/k_B$. The RDF plot reveals two sharp peaks and a diffusion pattern for both pure and plasticized glass samples. These peaks, representing the pair separation distances $r$ = 0.96a and $r$ = $2^{1/6}a$ where the FENE (intra-particle pair) and LJ (inter-particle pair) potentials reach their minimum values, respectively, and serve as indicators of particle interactions providing insights into the internal structure. The RDF plot indicates that the addition of diluent diminishes the number of intraparticle interactions (arising from chemical bonds) while increasing the number of interparticle (or Van der Waals) interactions due to the introduction of more short chains, resulting in a higher number of free chain-ends. Notably, the pair separation distance $r$ = $2^{1/6}a$, representing the minimum potential energy state of interparticle interactions, exceeds $r$ = $0.96a$, the distance for intraparticle interactions, leading to an increase in specific volume and free volume upon diluent addition. This augmentation facilitates greater molecular mobility compared to a pure sample, thereby contributing to the reduction in glass transition temperature $T_g$, as depicted in Figure \ref{fig:GlassTransition}e. 

Another significant factor that affects the glass transition temperature is the presence of entanglements. In the case of the pure polymer melt sample ($\phi$ = 0\%), as the temperature decreases, polymer chains begin to approach each other, leading to the formation of entanglements. These entanglements restrict the mobility of long chains and slow down segmental dynamics.

However, when diluent chains are added ($\phi$ > 0\%), the number of short chains in the system increases and host polymer chains decrease. Upon reducing the temperature of the bidisperse samples, these short chains freeze without forming entanglements. Consequently, only the remaining percent of host polymer chains participate in the formation of the entanglements; as a result, the average entanglement per chain <$Z$> in the bidisperse samples is lower compared to the pure sample $\phi$ = 0\% (See Figure \ref{fig:Z1entanglement}), allowing for greater molecular mobility. Consequently, this contributes to a reduction in the glass-transition temperature.

\subsection{Tensile-Test and Toughness Measurement}

\begin{figure}[H]
\centering
    \includegraphics[scale=0.9]{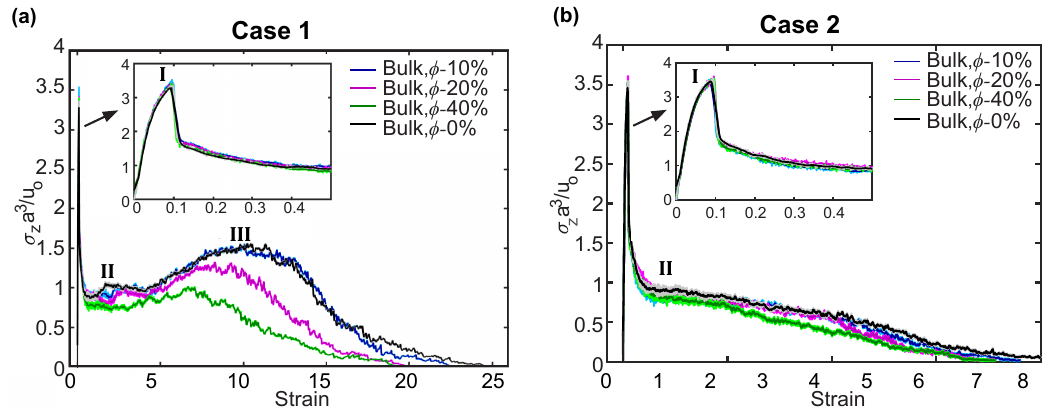}
    \caption{\textbf{Uniaxal tensile test of blended glassy polymer samples at $T$ = 0.3 $u_o/k_B$.} \textbf{a}, stress-strain response of Case 1 samples, \textbf{b} stress-strain response of Case 2 samples.}
    \label{fig:TensileTest}
\end{figure}

To ensure stable film formation and deformation-induced bonding, films must exhibit ductile characteristics and demonstrate bulk plastic deformation under stress. Therefore, we conducted a tensile test (See Method Section) to assess the stress-strain response of each polymer blend sample. Figures \ref{fig:TensileTest}a and b show the stress-strain response of both Case 1 and Case 2 samples under the uniaxial tensile test.

Both Case 1 and Case 2 samples exhibit ductility and show an elastic zone, a plastic zone, and a yield strength, which ensures a stable mechanical response\cite{zhang2021thickness}. However, only Case 1 samples (See Figure \ref{fig:TensileTest}a) show all three characteristic regimes: I, cavity nucleation; II, craze growth; and III, craze failure\cite{rottler2003growth}. By contrast, regime III (craze failure) is absent in Case 2 samples (See Figure \ref{fig:TensileTest}b), as short-chain samples failed before achieving uniform craze formation\cite{rottler2003growth}.

Further, we noted a reduction in craze growth with the addition of diluent chains, as evidenced by the downward shift of regime II in the stress-strain curves with increasing $\phi$, as illustrated in Figure \ref{fig:TensileTest}a. This reduction in craze growth can be solely attributed to the decrease in the number of host polymer chains $M_h$ that meet the condition $N_h>2N_e$, which is responsible for craze growth and craze failure\cite{rottler2003growth}. 

\begin{figure}[H]
\centering
    \includegraphics[scale=1.0]{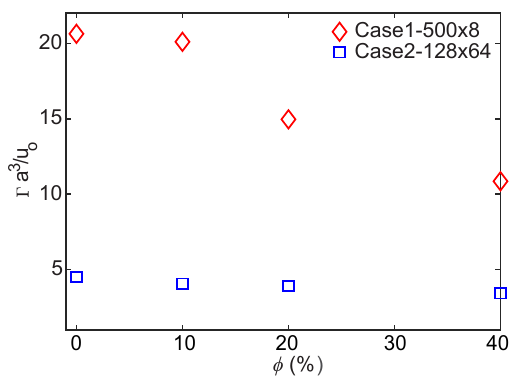}
    \caption{\textbf{Bidisperse polymer blend samples toughness as a function of diluent chain fraction $\phi$.}}
    \label{fig:toughness}
\end{figure}

We also quantify the toughness $\Gamma=\int\sigma_zd\epsilon$ of the bidisperse blend and homopolymer samples of Case 1 and Case 2 by integrating the stress-strain curves present in Figure \ref{fig:TensileTest}. Figure \ref{fig:toughness} shows that both Case 1 and Case 2 samples show a decrease in toughness as diluent concentration ($\phi$ > 0\%) increases.  

\subsection{Deformation-Induced Bonding and Debonding}

\subsubsection{Evolution of Stress and Normalized Entanglement Density in the Interface Region} 

\begin{figure}[H]
\centering
    \includegraphics[width=1.0\textwidth]{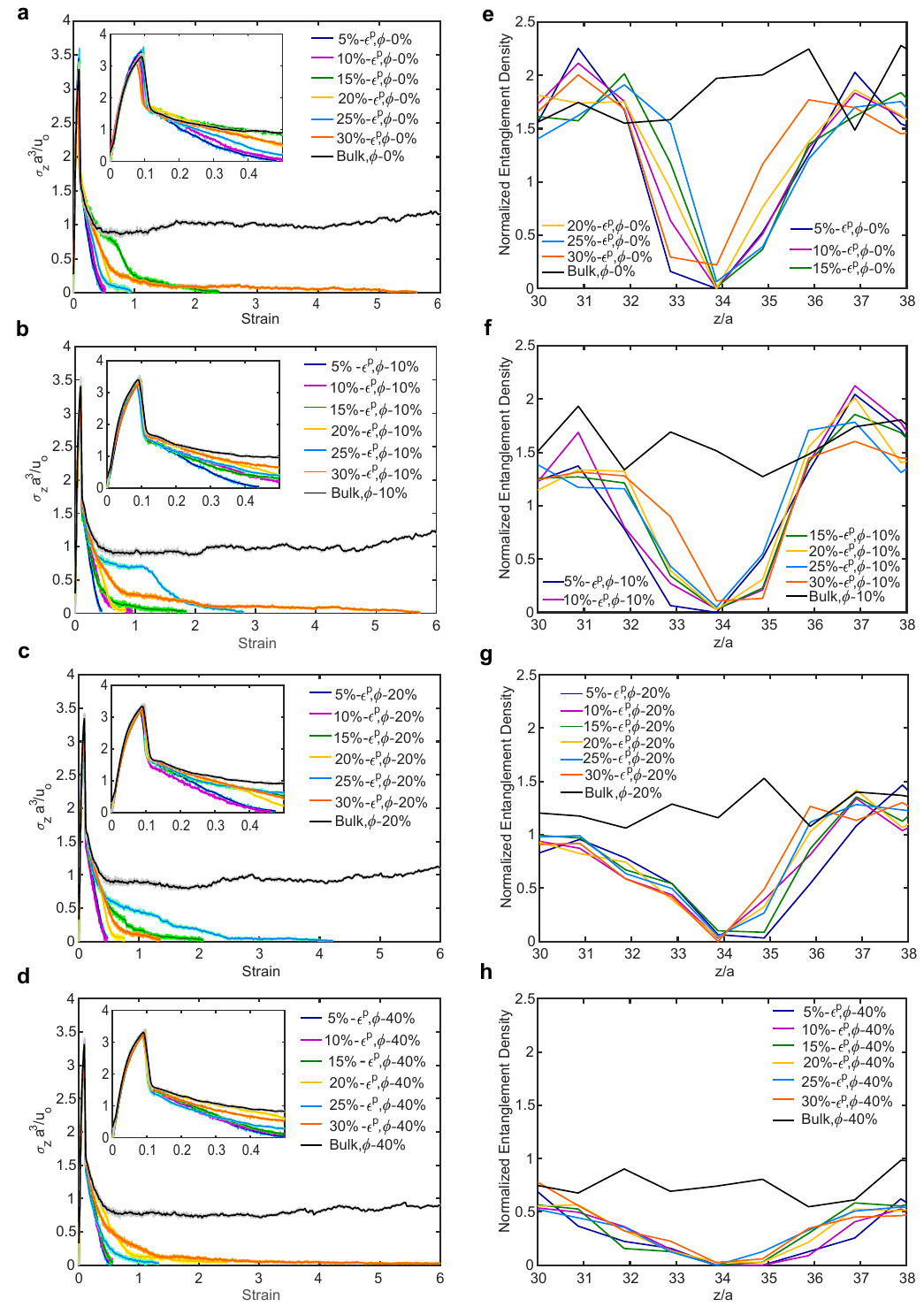}
    \caption{\textbf{Tensile test and normalized entanglement density analysis of bonded samples of Case 1.} \textbf{a, b, c, and d}, represent the stress-strain response of bulk bidisperse-samples bonded at 5\%, 10\%, 15\%, 20\%, 25\%, and 30\% plastic-strains. \textbf{e, f, g, and h}, represent the normalized entanglement density of bonded samples with respect to plastic strain.}
    \label{fig:Bonding}
\end{figure}

Figures \ref{fig:Bonding}a, b, c, and d depict the debonded stress-strain response of $\phi$ = 0\%, 10\%, 20\%, and 40\% diluent containing bulk samples of Case 1, bonded at $\epsilon^p$= 5\%, 10\%, 15\%, 20\%, 25\%, and 30\% plastic strains at $T$ = 0.3 $u_o/k_B$. The debonding stress-strain curves of both pure and bidisperse samples exhibit an elastic region followed by maximum elastic stresses ranging between 3.4 $\sigma_z a^3/u_o$ and 3.6 $\sigma_z a^3/u_o$. Subsequently a sharp drop is observed, indicating the onset of cavity nucleation \cite{rottler2003growth}.

All curves display a similar cavity nucleation regime (I) at different $\epsilon^p$, confirming that cavity nucleation is controlled by local internal structure and stress \cite{rottler2003growth}. However, stable craze growth (II) and craze failure (III) regimes are absent in these debonded curves, suggesting that the bonding strengths developed between samples are still below the bulk strength. This limitation could arise because the bonding experiments (plane strain compression test) for Case 1 samples were conducted at $T$ = 0.3 $u_o/k_B$, significantly below the glass transition temperature $T_g\approx0.45u_o/k_B$. Snapshots of strongly bonded samples from Case 1 and Case 2 during the tensile test at $T$ = 0.3 $u_o/k_B$ are presented in Supporting Figures \ref{fig:FractureCase1} and \ref{fig:FractureCase2}, respectively. These snapshots confirm the presence of chain pullout and chain scission modes of craze failure during the tensile test.
We also conducted an entanglement analysis to examine the growth of entanglements in the interface region. During the bonding experiments, plane strain conditions were applied in the y-direction, allowing the samples to expand in the x-direction. Therefore, we computed the normalized entanglement density, defined as the total number of entanglements in the $a$ unit thickness layer along the z-direction divided by the length of the bonded interface $L_x$ in the x-direction. Figures \ref{fig:Bonding}e, f, g, and h illustrate the normalized entanglement density plot as a function of the z-dimension. In the calculations of normalized entanglement density, we excluded all low-molecular-weight diluent chains ($N_d$ = 8), as mentioned earlier, due to their inability to withstand mechanical loads \cite{rottler2002cracks,bukowski2021load}. Consequently, the normalized entanglement density of both bulk and bonded samples in Case 1 reduces as the diluent chain fraction $\phi$ > 0\% increases. This trend is evident as we move from Figure \ref{fig:Bonding}e to Figures \ref{fig:Bonding}f, g, and h, respectively.

\subsubsection{Interfacial Fracture Energy}

To measure the interfacial fracture energy $G_I$ of bonded samples, we follow the established procedure outlined in previous simulations \cite{ge2013molecular,ge2014tensile}, which is applicable only for bonded samples that do not form stable craze during the uniaxial tensile test. In this approach, $G_I$ is considered equivalent to the interfacial tensile strength of the bonded samples. Then, we can quantify $G_I$ = $\int\sigma_zLz_od\epsilon$ of the bonded samples by integrating the stress-strain curves presented in Figures \ref{fig:Bonding}a, b, c, and d.

\begin{figure}[H]
\centering
    \includegraphics[width=1.0\textwidth]{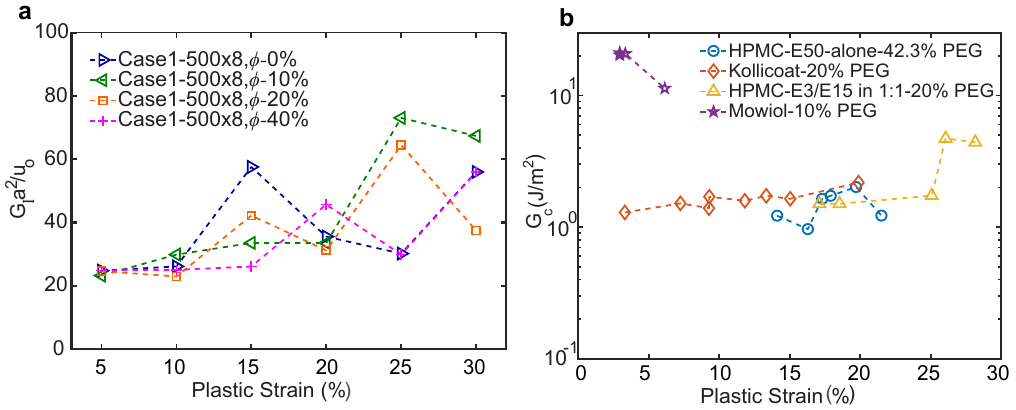}
    \caption{\textbf{Fracture energy of debonded samples.} \textbf{a}, Interfacial fracture energy $G_I a^2/u_o$ versus plastic strain plot for $\phi$ = 0\%, 10\%, 20\%, and 40\% bidisperse samples of Case 1. \textbf{b}, Fracture toughness $G_c$ ($J/m^2$) versus plastic strain plot for HPMC-E50-alone-42.3\% PEG, Kollicoat-20\% PEG, HPMC-E3/E15 in 1:1-20\% PEG, and Mowiol-10\% PEG taken from\cite{padhye2021deformation}.} 
    \label{fig:WorkofFracture}
\end{figure}

Figure \ref{fig:WorkofFracture}a illustrates the interfacial fracture energy $G_I$ of $\phi$ = 0\%, 10\%, 20\%, and 40\% bidisperse bulk samples of Case 1, bonded at $\epsilon^p$= 5\%, 10\%, 15\%, 20\%, 25\%, and 30\% plastic strains at $T$ = 0.3 $u_o/k_B$. The $G_I$ curves obtained from simulations (See Figure \ref{fig:WorkofFracture}a) exhibit a non-monotonic trend, which is consistent with experimental results\cite{padhye2021deformation} (See Figure \ref{fig:WorkofFracture}b) and previous simulations\cite{padhye2022dilatational} (Note: In previous simulations, we defined bonding strength in terms of work of fracture $W_f$=$\int\sigma_zd\epsilon$). The reported $G_I$ values of debonded bidisperse-samples, ranging between 23.22 $u_o/a^2$ and 73.05 $u_o/a^2$, are significantly larger than the free energy associated with two interfaces ($G_I>$ 2$\gamma$), where $\gamma\sim$ 1$u_o/a^2$ is the interfacial energy of the polymer-vacuum interface\cite{gersappe1999polymer,baljon2001simulations}. This is only possible through opposite-side entanglement formation at the interface during deformation-induced bonding.

The simulation results also indicate that the $\phi$ = 10\% bidisperse samples exhibit the potential to achieve the highest interfacial fracture energy, with $(G_I)_{max}$ = 73.05 $u_o/a^2$, followed by $\phi$ = 20\% samples ($(G_I)_{max}$ = 64.47 $u_o/a^2$), $\phi$ = 0\% ($(G_I)_{max}$ = 57.55 $u_o/a^2$), and finally $\phi$ = 40\% samples ($(G_I)_{max}$ = 55.87 $u_o/a^2$). This trend aligns qualitatively with experimental observations\cite{padhye2021deformation}, wherein the Mowiol-10\% PEG samples exhibited the highest fracture toughness $(G_c)_{max}$ = 21 $J/m^2$, followed by HPMC-E3/E15 in 1:1-20\% PEG samples ($(G_c)_{max}$ = 4.709 $J/m^2$), Kollicoat-20\% PEG samples ($(G_c)_{max}$ = 2.17 $J/m^2$), and then HPMC-E50-alone-42.3\% PEG ($(G_c)_{max}$ = 2.01 $J/m^2$). It is especially important to note here that the host chain polymer type in those experiments differed while the diluent chains were of the same composition, thus emphasizing the critical role of chain length and bidisperse volume fraction which predominantly impacts number of chain ends and entanglement topology, rather than being dominated by host chain structural differences.   

\subsubsection{Molecular mobility, chain-ends and entanglements}

\begin{figure}[H]
\centering
    \includegraphics[width=1.0\textwidth]{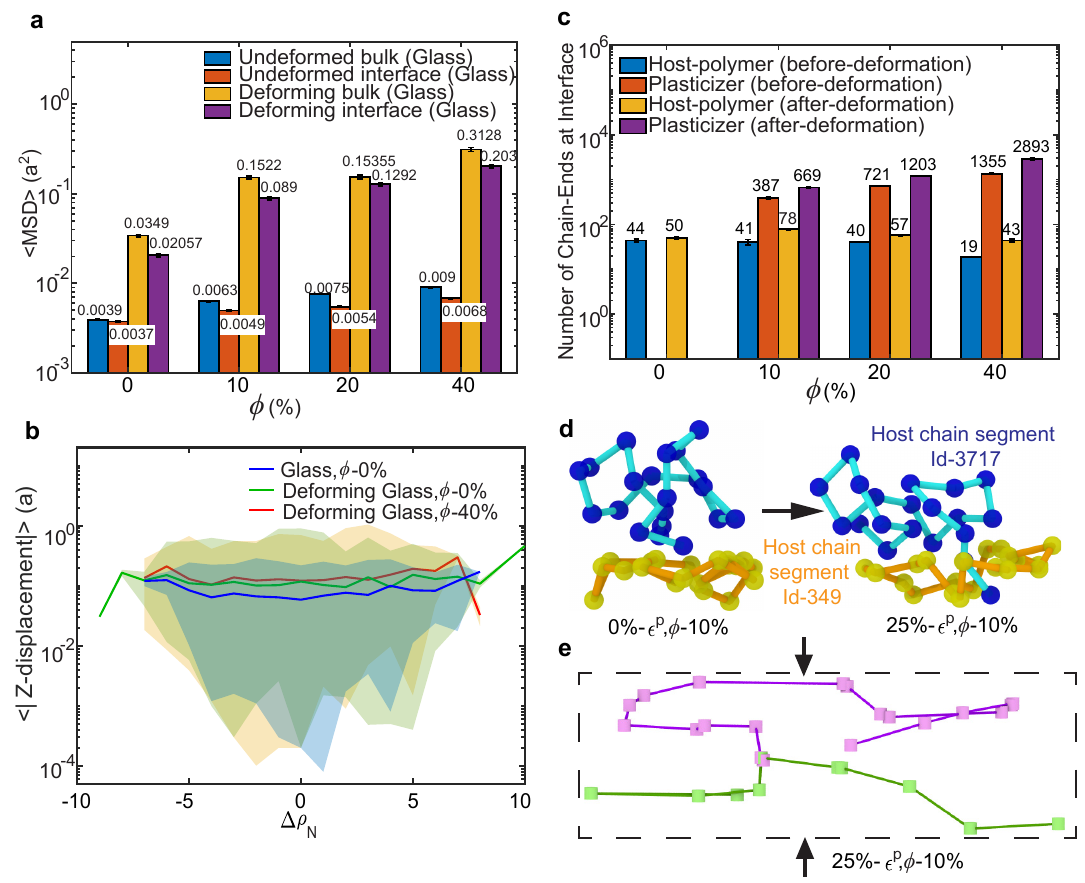}
    \caption{\textbf{Molecular mobility, chain-ends, and entanglement formation.} \textbf{a}, Bead displacement in undeformed and deforming glass with respect to diluent chain concentration $\phi$. \textbf{b}, comparison of chain-end absolute Z-displacements with respect to gradient in local number density. \textbf{c}, number of chain-ends at the interfacial region before and after deformation, corresponding to diluent chain concentration $\phi$. \textbf{d} entanglement formation between chain Id-349 and Id-3717 at 25\%-$\epsilon^p$, $\phi$-10\%. \textbf{e}, demonstration of primitive-paths of opposite-side chain-Ids (349, 3717) at 25\%-$\epsilon^p$, $\phi$-10\%.} 
    \label{fig:Mobility}
\end{figure}

To examine the influence of diluent concentration $\phi$ on the mobility of beads during plastic deformation and in the glassy state within the bulk and interfacial regions, we calculated the time-average mean square displacement <MSD> of beads in these regions at intervals of 50 $\tau$ (See Supplementary subsection 5.1, and Figure \ref{fig:MSD2}). The bar plot in Figure \ref{fig:Mobility}a indicates that as the diluent concentration $\phi$ increases, <MSD> also increases, doubling in the glassy state at $\phi$ = 40\%. In the deformation state, it exhibits nearly two orders of magnitude mobility enhancement in these regions compared to the pure glass ($\phi$ = 0\%). 

However, mobility does not alone fully represent the cause of strong bonding. For strong bonding, opposite-side entanglements are required, which is only possible through polymer chain interdiffusion. In our prior research \cite{padhye2022dilatational}, we established that the primary mechanism underlying DIB is dilatation (densification)-based interdiffusion \cite{padhye2022dilatational,argon1979plastic}, contrasting with reptation-based interdiffusion \cite{de1971reptation}, where polymer chains navigate within an imaginary tube \cite{doi1988theory} formed by the topological constraints of surrounding polymer chains. However, in dilatation-based interdiffusion, local sites, or shear transformation zones\cite{argon2013physics} (STZs), diffuse in random directions by local bead rearrangement through shear localization\cite{argon1979plastic} during plastic deformation \cite{padhye2022dilatational,argon2013physics}. Consequently, both chain-end and non-chain-end beads penetrate to the opposite side of the interface. This results in some chain-ends contributing to new entanglement formation\cite{padhye2022dilatational}, while non-chain ends contribute to kink formation\cite{argon1973theory,padhye2022dilatational} at the interface region.

To correlate the dilatation (densification) with interdiffusion, we computed the absolute Z-displacement of chain-ends at the interface region with respect to the temporal density gradient $\Delta\rho_N$\cite{padhye2022dilatational} for the pure glass (at $\phi$ = 0\%), pure deforming glass (at $\phi$ = 0\%), and plasticized-deforming glass (at $\phi$ = 40\%) by following our former procedure\cite{padhye2022dilatational} (See Figure \ref{fig:Mobility}b, and Supplementary Figure \ref{fig:Dilatation}). Here, $\Delta\rho_N$ represents the difference in local number density $\rho_N$ of the molecular site at each 50$\tau$ time step\cite{padhye2022dilatational}. The local number density $\rho_N$ is determined by encompassing a 2.5$a$ unit radius sphere around each molecular bead site, with the cumulative number of beads within that sphere defining the local number density of the respective bead. If the $\Delta\rho_N$ value of a local bead site is negative, that specific event is termed dilatation ($\Delta\rho_N$ < 0). Conversely, if it is positive, the event is referred to as densification ($\Delta\rho_N$ > 0) in our calculations. Figure \ref{fig:Mobility}b illustrates that the Z-displacement of chain-ends increases in deforming glasses ($\phi$ = 0\% and 40\%) at the interface boundary compared to the pure glass ($\phi$ = 0\%), indicating chain-end interdiffusion. We also computed the depth of penetration of interfacial regions for bonded samples of different concentrations $\phi$ with respect to plastic deformation (See Supplementary Figure \ref{fig:Depth}). 

Since only host polymer chain entanglements in Case 1 can bear the mechanical load \cite{bukowski2021load}, it becomes crucial to quantify the number of chain-ends of host polymer chains that are present at the interfacial region during deformation. Figure \ref{fig:Mobility}c illustrates the count of chain-ends belonging to the host polymer and to diluent polymer within the interface region both before and after deformation, with respect to diluent chain concentration $\phi$. The reported data of two sets indicates that the 10\% bidisperse sample exhibits the highest number of host polymer chain-ends (78$\pm$5) within the interface region after deformation, compared to the 0\% (50$\pm$5), 20\% (57$\pm$2), and 40\% (43$\pm$4) bidisperse samples, respectively.

The drastic enhancement in host polymer chain-ends at the interfacial region in the 10\% and 20\% bidisperse samples (compared to the pure glass, $\phi$ = 0 \%) arise from a combination of three factors: 1) enhanced mobility, 2) lower average number of entanglements per chain (<Z>) with respect to pure glass $\phi$ = 0\%, and 3) a sufficient quantity of host polymer chains ($M_h$= 450 and 400, respectively). However, in the case of the 40\% bidisperse samples, the third factor, with a lower number of host polymer chains ($M_h$= 300), is considerably diminished. Consequently, the number of host polymer chain-ends available in the interfacial region before and after deformation decreases.

This suggests that the 10\% bidisperse samples have a greater likelihood of forming opposite-site host polymer chain entanglements compared to the 0\% host homopolymer and the 20\%, and 40\% bidisperse samples, respectively. This aspect could serve as a pivotal factor contributing to the potential attainment of higher interfacial fracture energy ($G_{I,max}$) by the 10\% bidisperse samples as opposed to the other samples during debonding, as depicted in Figures \ref{fig:WorkofFracture}a and b. Figure \ref{fig:Mobility}d shows the entanglement formation between two opposite-side host polymer chain segments (Ids 349, 3717) during plane strain compression (See Supplementary Video 1). Figure \ref{fig:Mobility}e is showing the primitive path\cite{everaers2004rheology} (i.e., geometrically constructed topologically constrained path between chain-ends with no chain crossability) of these polymer chains (Ids 349, 3717 at 25\%-$\epsilon^p$, $\phi$-10\%) drawn by using the Z1+ algorithm\cite{kroger2023z1+}. Each cylinder in Figure \ref{fig:Mobility}e represents a kink or topological constraint formed by the surrounding chains. The overlapping of the purple and green cylinders confirms that the entanglement is forming between chains Id-349 and Id-3717.

\subsection{Strong bonding}

\begin{figure}[H]
\centering
    \includegraphics[scale=1.0]{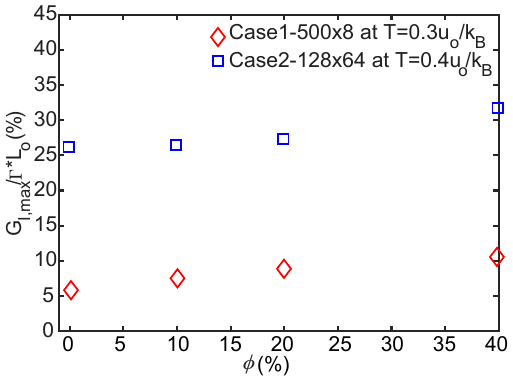}
    \caption{\textbf{Maximum strength of bonded samples with respect to bulk.}} 
    \label{fig:Strength}
\end{figure}

\begin{figure}[H]
\centering
    \includegraphics[width=1.0\textwidth]{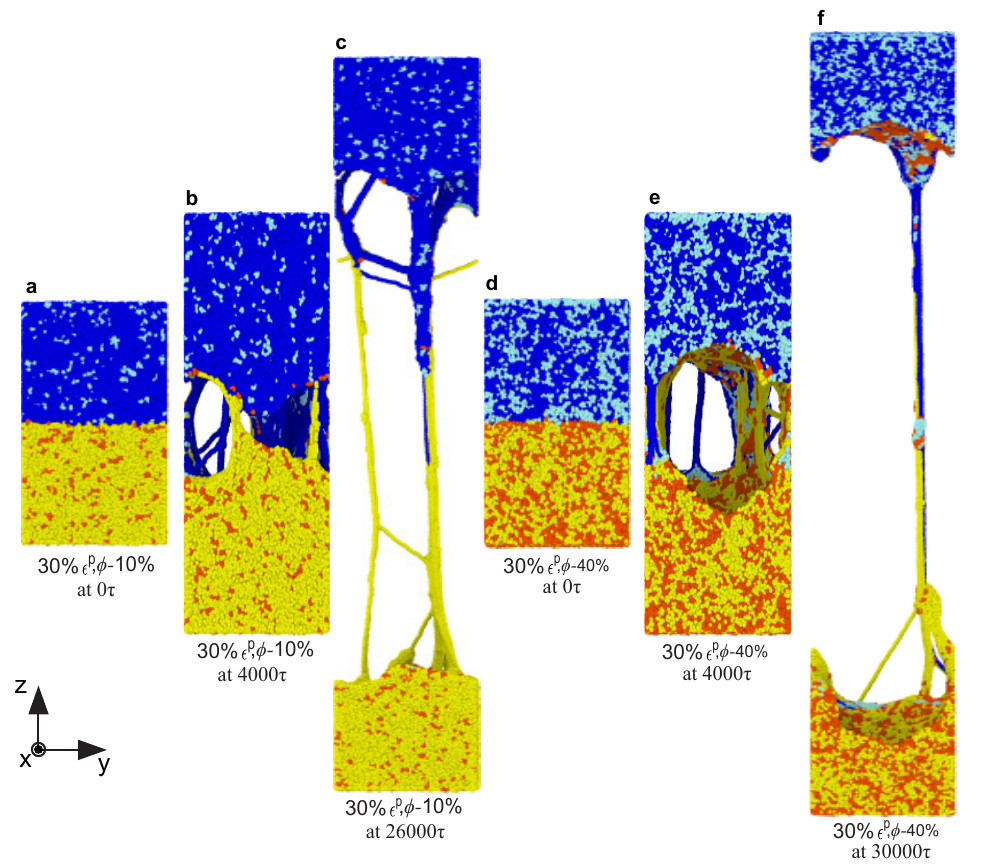}
    \caption{\textbf{Snapshot of failure of Case 1 Samples during tensile tests.} Here, dark blue and yellow colors represent host polymer chains, while sky blue and red colors represent diluent polymer chains.} 
    \label{fig:FractureCase1}
\end{figure}

\begin{figure}[H]
\centering
    \includegraphics[width=1.0\textwidth]{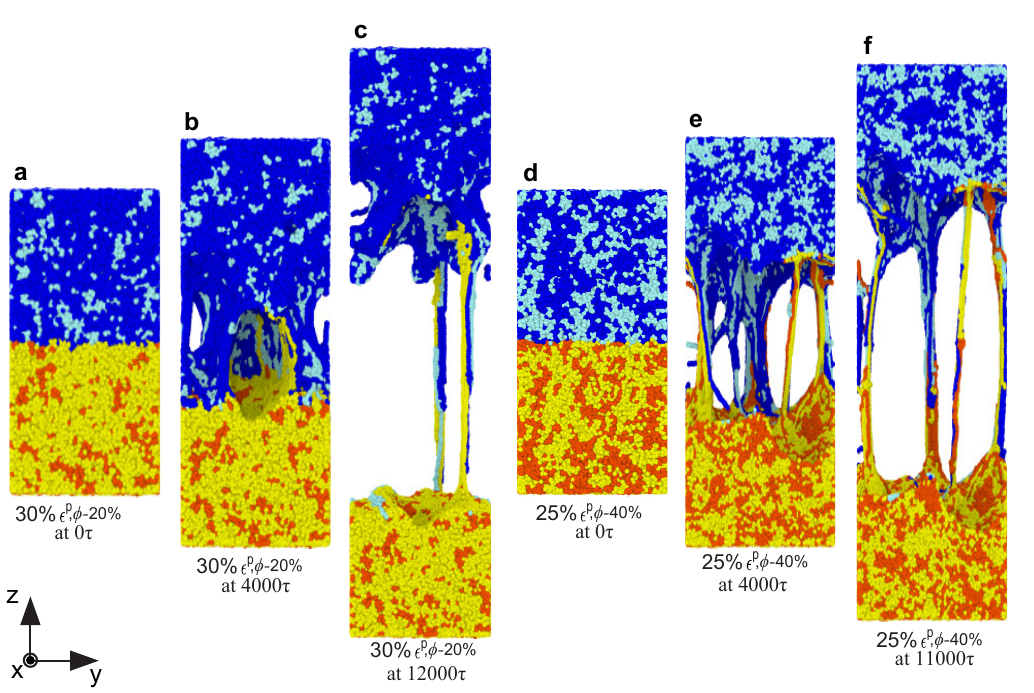}
    \caption{\textbf{Snapshot of failure of Case 2 Samples during tensile tests.} Here, dark blue and yellow colors represent host polymer chains, while sky blue and red colors represent diluent polymer chains.} 
    \label{fig:FractureCase2}
\end{figure}

In Section 3.4, we presented the analysis of deformation-induced bonding of Case 1 samples at temperature $T$ = 0.3 $u_o/k_B$. Figure \ref{fig:Strength} illustrates that the bonding strength of these samples varies between 6\% to 11\% relative to the diluent concentration $\phi$, compared to their corresponding bulk strength ($\Gamma$). Notably, the bonding temperature $T$ = 0.3 $u_o/k_B$ is significantly lower than the bulk-glass transition temperature $T_g^b \approx$ 0.45 $u_o/k_B$, which could be a primary factor contributing to the observed weak bonding with respect to bulk.

In response, we designed Case 2 samples with lower host molecular weight, featuring a moderately entangled host polymer ($N_h$=128, where $N_h\ge2N_e$) paired with a weakly entangled diluent ($N_d$= 64, with kink $Z>1$). Bonding experiments were conducted at $T$ = 0.4 $u_o/k_B$, still below the bulk $T_g\approx$ 0.443 $u_o/k_B$ yet with less of a delta compared to Case 1. This choice of material and process conditions was motivated by the analysis of Case 1 sample data and the hypothesis that reducing the molecular weight increases the number of chain-ends on the free surface. This, in turn, facilitates interdiffusion and entanglement formation, ultimately leading to stronger bonding. However, it is crucial that the molecular weight of the host polymer chain be at least twice that of the molecular weight of the entanglement ($M_h\ge2M_e$ or $N_h\ge2N_e$) for stable film formation (stable craze formation\cite{rottler2003growth}).

Our observations revealed that the low molecular weight diluent polymer chains ($N_p=8$, with kink $Z\le1$) in Case 1 samples also form entanglements during bonding experiments (See Supplementary Figure \ref{fig:Entanglementsup}, and Table \ref{tab:Entanglement}), yet they do not contribute as load-bearing entanglements\cite{rottler2002cracks}. This served as our motivation for adjusting the diluent polymer chain length ($N_d=64$, with kink $Z>$1) in Case 2 samples. Literature suggests that increasing the bonding temperature with respect to the polymer glass enhances its ductility, and deformation amplifies existing high mobility to facilitate rapid bonding; therefore, we increased the bonding temperature condition from $T$ = 0.3 $u_o/k_B$ to $T$ = 0.4 $u_o/k_B$. 

Figure \ref{fig:Strength} finally shows that the bonding strength of Case 2 samples varies between 26\% to 32\% relative to the diluent polymer concentration $\phi$, compared to their corresponding bulk strength ($\Gamma$). This marks a significant improvement over the bonding strength observed in Case 1 samples.         

\section{CONCLUSIONS}

In summary, we have examined thermophysical and mechanical properties such as glass transition temperature, internal structure (RDF plot), stress-strain response, failure mode, and toughness of bidisperse blends via variation in diluent polymer concentration ($\phi$) of chemically identical homopolymers with different molecular weights using MD simulations. The mechanical properties of Case 1 samples undergo profound transformations in contrast to Case 2 samples corresponding to $\phi$. This disparity arises due to the integration of a low-molecular-weight diluent polymer ($N_d$ = 8, $Z\le1$) with the host polymer exclusively in Case 1 samples.

We also critically examine the role of the lower-molecular-weight diluent ($N_d$ = 8, $Z\le1$) in the deformation-induced bonding of bidisperse blend samples at $T=0.3 u_o/k_B$, i.e., below both surface and bulk glass transition temperatures, $T_g^s$ and $T_g^b$. We found that the addition of diluent enhanced the molecular mobility of the polymer system, which was further amplified (by almost two orders of magnitude compared to the pure glass state) during deformation-induced bonding. Results suggest that a relatively low fraction of diluent polymer concentration ($\phi\le$ 20\% in a polymeric system significantly enhances the number of host polymer chain-ends at the interface region compared to a pure glass ($\phi$ = 0\%) system during deformation, which ultimately improves the likelihood of opposite-side host polymer chain entanglement formation. As a result, the $\phi$ = 0\% and 20\% bidisperse samples of Case 1 have the potential to achieve the maximum interfacial fracture energy $G_{I,max}$ compared to the 0\% and 40\% samples of Case 1. 

The Case 2 results show that by optimizing material and process conditions, significant improvement in bonding strength (almost one third of bulk strength) can be achieved even at temperatures below the bulk glass transition temperature $T_g^b$. However, in Case 2, there are multiple factors involved, such as diluent chain length ($N_d$ = 64, $Z$ > 1), which can also contribute to bonding strength as these longer chains can bear mechanical load\cite{rottler2002cracks}; bonding experiment temperature ($T$ = 0.4 $u_o/k_B\ge T_g^s$), etc.; therefore, no direct correlation was found between maximum interfacial fracture energy ($G_{I,max}$) and diluent chain concentration ($\phi$) in Case 2. However, the maximum bonding strength to bulk strength ratio ($G_{I,max}/(\Gamma*L_o$)) in both Case 1 and Case 2 monotonically increased with respect to diluent polymer concentration ($\phi$).

While our focus has predominantly centered on elucidating the impact of molecular weight and volume fraction ($\phi$) of bidisperse blends of chemically identical homopolymers on deformation-induced bonding, it is worth noting that several other factors, including bonding temperature, strain rate\cite{padhye2022dilatational} (previous simulation work was performed at higher deformation rate as plastic deformation was accomplished in 2000 $\tau$ time step), and the interplay of chemically diverse macromolecules, can also exert influence on bonding strength. These intriguing avenues for exploration are left for future investigation, promising further transformative insights.

\section{Supplementary}

\textbf{\begin{figure}[H]
\centering
    \includegraphics[scale=0.4]{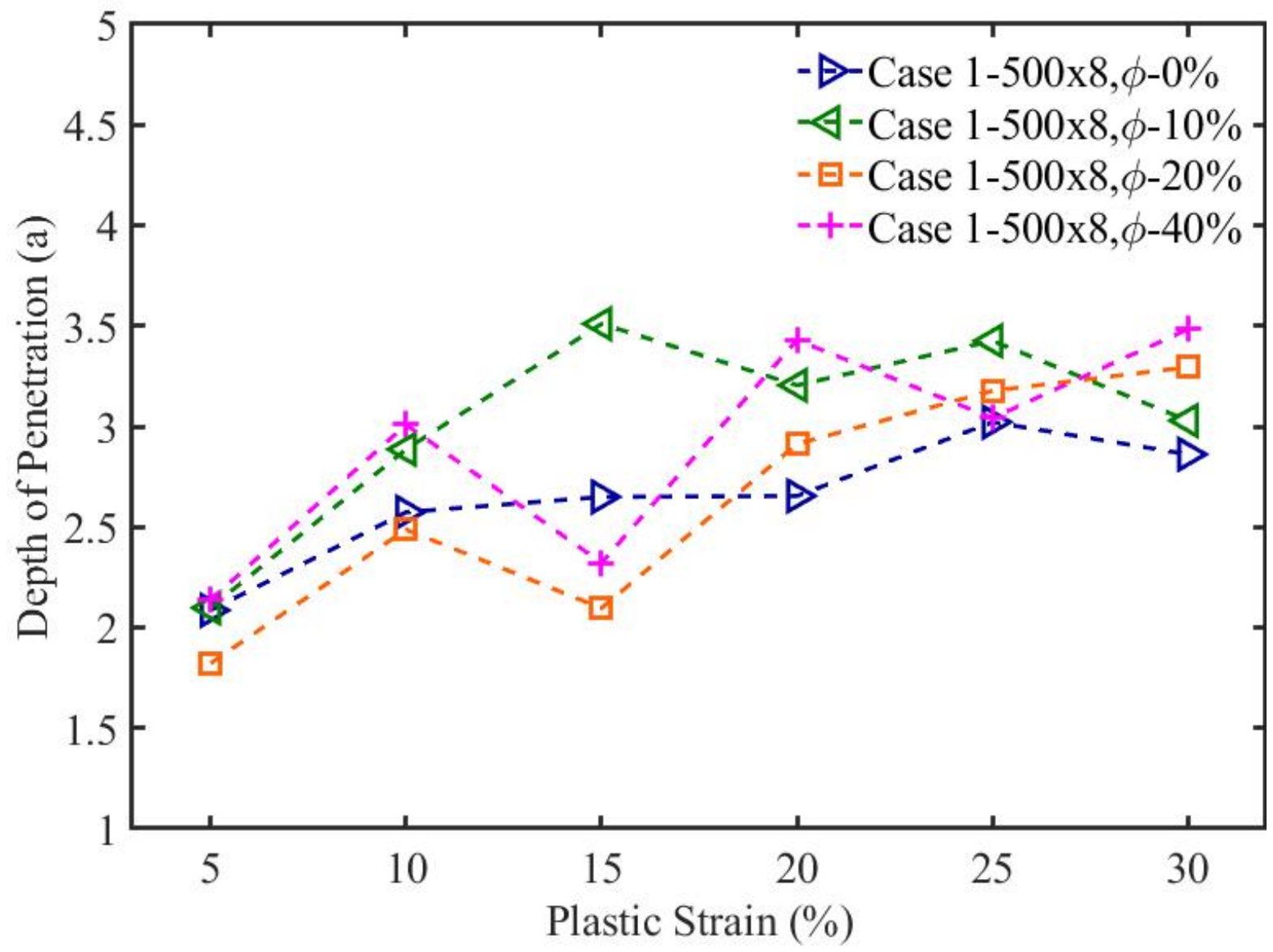}
    \caption{\textbf{Depth of penetration of the interfacial region of Case 1 samples with respect to plastic deformation}} 
    \label{fig:Depth}
\end{figure}}

\subsection{<MSD> calculation of beads in the bulk and interfacial regions of undeformed and deforming glass samples.}

\begin{figure}[H]
\centering
    \includegraphics[scale=1.0]{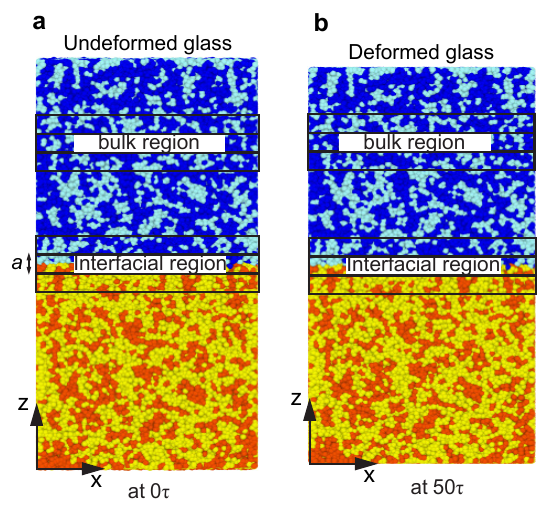}
    \caption{\textbf{Schematic diagram to define bulk and interfacial region in undeformed and deformed glass samples}} 
    \label{fig:MSD2}
\end{figure}

To compute the bead mobility in both the bulk and interfacial regions of undeformed and deforming glass samples, we divided the samples into layers of thickness $a$ along the z-axis, as illustrated in Figure \ref{fig:MSD2}. Subsequently, we calculated the mean-square displacement of beads within each layer at a time interval of $\Delta t = t^{i+1} - t^{i} = 50\tau$ (total time of deformation is around 4000$\tau$). The choice of time interval was based on previous simulations, ensuring adequate characterization of mobility \cite{morita2006study}. The MSD of the beads in the layer $l$ is defined by
\begin{equation}
MSD_l(\Delta t)=\dfrac{\sum_{i=1}^{}\sum_{\textrm{n in layer }l}^{}[r_n(t^{i+1})-r_n(t^i)]^2}{\sum_{i=1}^{}\sum_{\textrm{n in layer }l}^{}1}    
\end{equation}
where $r_n$ is a position vector of bead n at time step i. The width of the interfacial region, as determined from the depth of penetration plot shown in Figure \ref{fig:Depth}, is approximately 3-4a units. The mean square displacement of the interfacial region, denoted as $<MSD^I>$, is calculated as the average $MSD_l(\Delta t)$ across three layers adjacent to the interface (See Figure \ref{fig:MSD2}a). Similarly, the mean square displacement of the bulk region, also denoted as $<MSD^b>$, is determined as the average across three layers near the centre (See Figure \ref{fig:MSD2}a).

\subsection{List of opposite side entanglements} 

\begin{figure}[H]
\centering
    \includegraphics[width=1.0\textwidth]{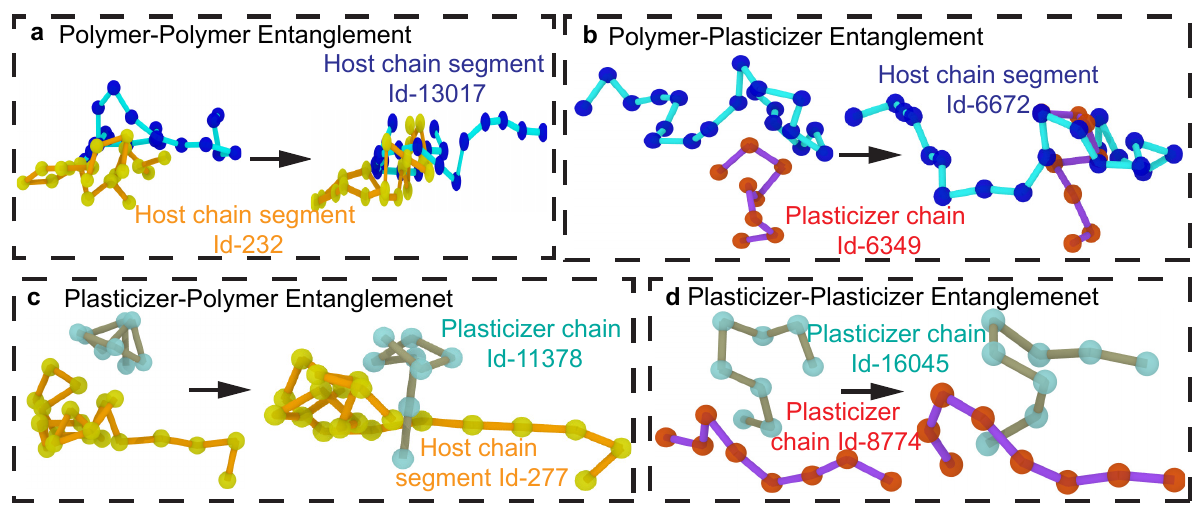}
    \caption{\textbf{Opposite-side entanglements formation during bonding in Case 1 samples at T=0.3 $u_o/k_B$}}
    \label{fig:Entanglementsup}
\end{figure}

Table \ref{tab:Entanglement} and Figure \ref{fig:Entanglementsup} illustrate various opposite-side entanglements that arise during bonding experiments, encompassing interactions between diluent-diluent, diluent-host polymer, and host-host polymer. It is important to note that while these examples are provided, they represent only a subset of the entanglements occurring during deformation-induced bonding; the actual number is much higher. The Z1+ algorithm facilitates quantification of kinks formed by each chain, although it does not directly attribute these kinks to specific entanglement pairs. The data presented in the table stems from the post-processing of Z1+ algorithm results. Despite this, there remains a challenge in accurately correlating type a, type c, type d, and type e kinks (See Figure 15 in reference\cite{karayiannis2009combined}) with their corresponding entanglement pairs.

\begin{table}[H]
    \caption{\textbf{Partial list of entanglements formed in Case 1 samples during bonding}}
    \centering
    \begin{tabular}{|c|c|c|c|c|c|c|}
    \hline
       Chain Id & Chain type & Kink & Location& Coordinate x& Coordinate y& Coordinate z\\ \hline
       \multicolumn{7}{|l|}{Case 1, $\phi$-10\% Entanglement IDs} \\\cline{1-7}
       25   & 1 & 1 & 267.17 & 20.90353 & 6.274582 & 32.08831 \\
       4991 & 4 & 1 & 6.44   & 20.90509 & 6.27717  & 32.0895 \\
        \hline
    3057 & 2 & 1 & 5.08 & 16.37256 & -11.1956 & 32.1463  \\
    5872 & 4 & 1 & 2.5  & 16.37004 & -11.1958 & 32.14633 \\
 \hline
203  & 1 & 1 & 455.67 & -13.2171 & 25.52949 & 29.72076 \\
6563 & 4 & 1 & 2.17   & -13.2171 & 25.53084 & 29.72137 \\       
 \hline  
332  & 1 & 1 & 249 & -13.9649 & 25.4677  & 30.31989 \\
6563 & 4 & 1 & 1   & -13.1916 & 25.55798 & 29.47961 \\
\hline
173  & 1 & 1 & 159.33 & -7.99275 & 6.546332 & 36.51674 \\
5833 & 4 & 1 & 6.94   & -7.98967 & 6.546182 & 36.5169  \\ 
\hline
349  & 1 & 1 & 331.11 & 32.91941 & -15.4897 & 29.48315 \\
3717 & 3 & 1 & 5.5    & 32.90046 & -15.489  & 29.48816 \\
\hline
3631 & 3 & 1 & 173.5  & -9.03315 & 4.718799 & 25.74687 \\
2271 & 2 & 1 & 7      & -9.03359 & 4.71712  & 25.74762 \\
\hline
\multicolumn{7}{|l|}{Case 1, $\phi$-40\% Entanglement IDs} \\\cline{1-7}
16045 & 4 & 1 & 3.5    & 26.31138 & 23.20113 & 29.63333 \\
8774  & 2 & 1 & 2.5    & 26.30693 & 23.20046 & 29.6341  \\
\hline
21483 & 4 & 1 & 2.28   & 35.48817 & 19.93098 & 30.17509 \\
9896  & 2 & 1 & 6.5    & 35.4874  & 19.92999 & 30.17577 \\
\hline
4982  & 2 & 1 & 3      & -20.5488 & 27.98679 & 31.17938 \\
15308 & 4 & 1 & 4      & -20.55   & 27.98486 & 31.18334 \\
\hline
1704  & 2 & 1 & 2      & -37.5291 & 26.77819 & 25.38959 \\
15733 & 4 & 1 & 2.5    & -37.5309 & 26.77532 & 25.39102 \\
\hline
1563  & 2 & 1 & 4      & 7.656783 & -6.63307 & 25.53846 \\
17543 & 4 & 1 & 2.5    & 7.658597 & -6.63195 & 25.53994 \\
\hline
232   & 1 & 1 & 496.5  & 54.51922 & 0.552938 & 25.72968 \\
13017 & 3 & 1 & 471.67 & -24.2764 & 0.553913 & 25.73009 \\
\hline
8215  & 2 & 1 & 5.75   & 22.88304 & -23.0491 & 23.61559 \\
20989 & 4 & 1 & 4      & 22.88102 & -23.0502 & 23.61838 \\
\hline
1563  & 2 & 1 & 4.67   & 7.443402 & -7.15967 & 23.66583 \\
17543 & 4 & 1 & 4      & 7.442435 & -7.16001 & 23.66679 \\
\hline
16045 & 4 & 1 & 1      & 22.48094 & 23.40822 & 23.06329 \\
8774  & 2 & 1 & 5.5    & 22.79555 & 23.42125 & 23.28502 \\
\hline
    \end{tabular}
    \label{tab:Entanglement}
\end{table}

\subsection{RDF calculations of bonded samples}

The radial distribution function, denoted as $g(r)$, provides insights into the probability of finding a bead (or atom) within a given distance shell, typically measured from a reference bead or atom. It not only characterizes the spatial distribution of particles but also offers clues about their packing density within the system. Constructing an RDF involves selecting a bead within the system and encircling it with a series of concentric spherical shells, each spaced at a fixed distance interval, denoted as $\Delta r$. Through LAMMPS simulations, data is collected by tallying the number of beads located within each shell. Subsequently, this count is normalized by dividing it by the volume of each shell and the average bead density across the system. Mathematically the formula is:
                            
$$g(r)=n(r)/(\rho_{avg}*4\pi*r^2*\Delta r)$$

where $g(r)$ is radial distribution function, $n(r)$ is average number of beads in each shell, $\rho_{avg}=N/V_{box}$, $r$ is the radius of shell, and $\Delta r$ is distance between two nearest shell; it is defined as $\Delta r=r_{cutoff}/h$. where $r_{cutoff}=3.5a$ and $h=200$, which is number of histogram bins or spherical shells.
LAMMPS simulations provide different states' data (solid, melt, and deformed samples) which are further used to construct RDF plots by using OVITO commercial software.

\begin{figure}[H]
\centering
    \includegraphics[width=1.0\textwidth]{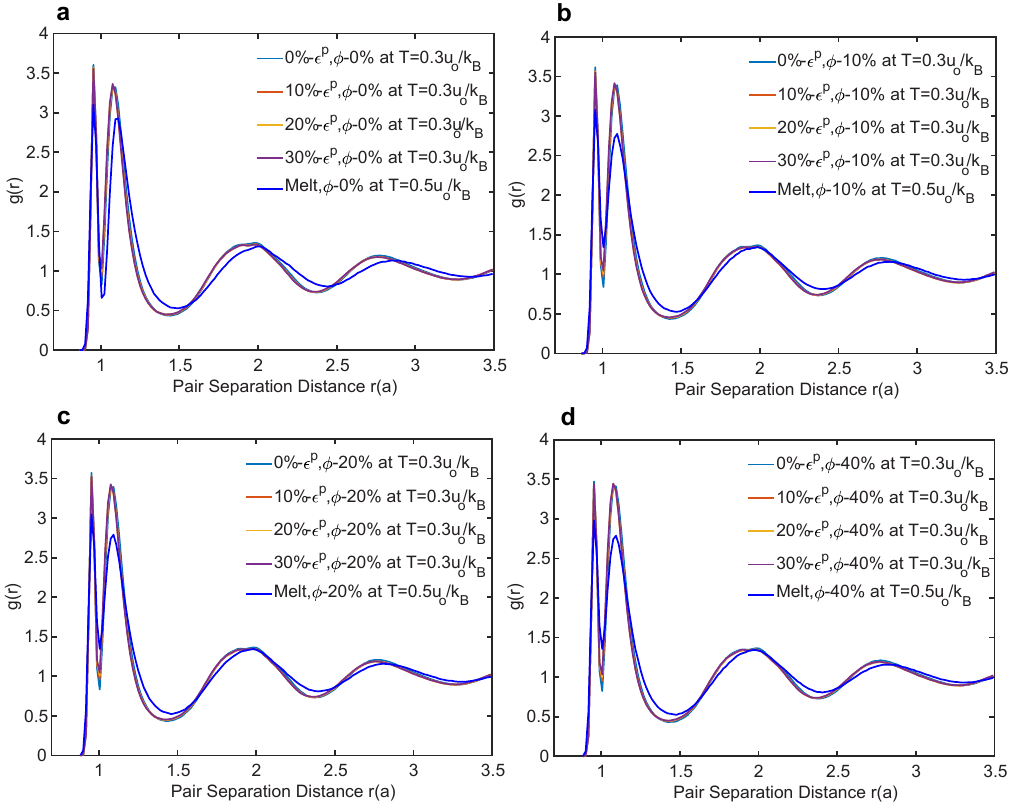}
    \caption{\textbf{RDF plots comparing DIB-bonded samples at 10\%, 20\%, and 30\%-$\epsilon^p$ with undeformed 0\%-$\epsilon^p$ glass and polymer melt samples of Case 1 at T=0.3 $u_o/k_B$.}} 
    \label{fig:RDFdeform}
\end{figure}

\subsection{Dilatation Plasticity}

The chain-end motion of blended samples was computed with respect to the gradient in local number density by following the standard procedure mentioned in the supplementary reference\cite{padhye2022dilatational}. 
\begin{figure}[H]
\centering
    \includegraphics[width=1.0\textwidth]{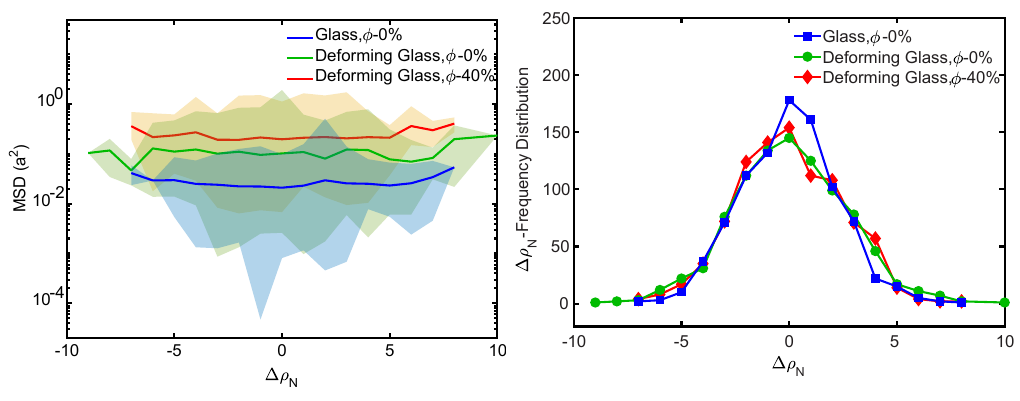}
    \caption{\textbf{Chain-end motion of Case 1 samples.} \textbf{a}, Comparison of chain-end displacements with respect to gradient in local number density and \textbf{b}, frequency distribution of gradient in local number density.} 
    \label{fig:Dilatation}
\end{figure}

\section{Extra Data}

\begin{figure}[H]
\centering
    \includegraphics[scale=1.0]{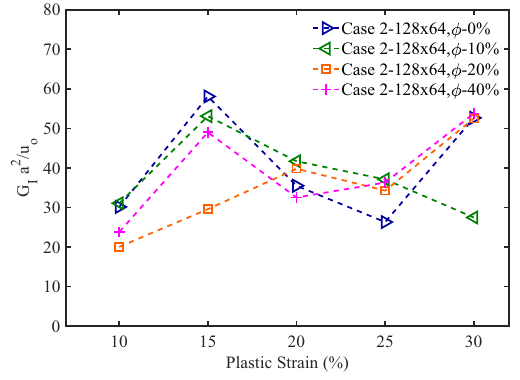}
    \caption{\textbf{Fracture energy of debonded samples of Case 2}}
    \label{fig:Fracture2}
\end{figure}

\begin{figure}[H]
\centering
    \includegraphics[width=1.0\textwidth]{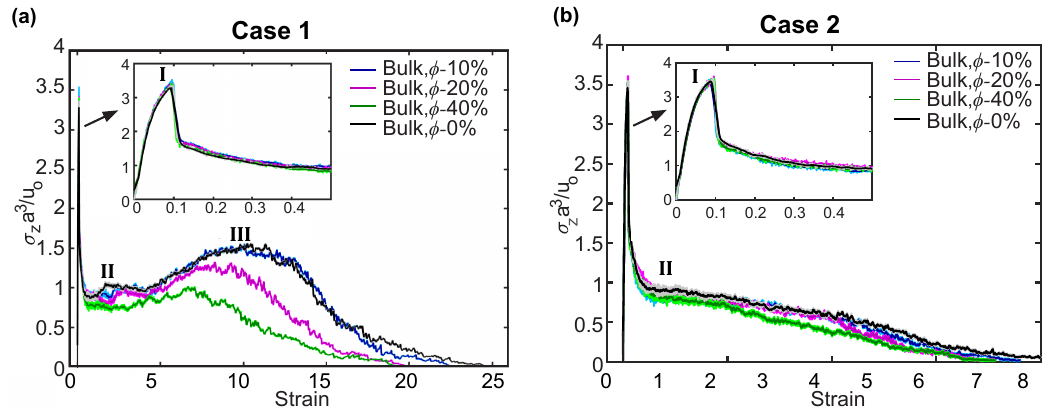}
    \caption{\textbf{Tensile test and normalized entanglement density analysis of bonded samples of Case 2.} \textbf{a, b, c, and d}, represent the stress-strain response of bulk bidisperse-samples bonded at 10\%, 15\%, 20\%, 25\%, and 30\% plastic-strains.}
    \label{fig:TensileTestplot}
\end{figure}

\begin{figure}[H]
\centering
    \includegraphics[scale=1.0]{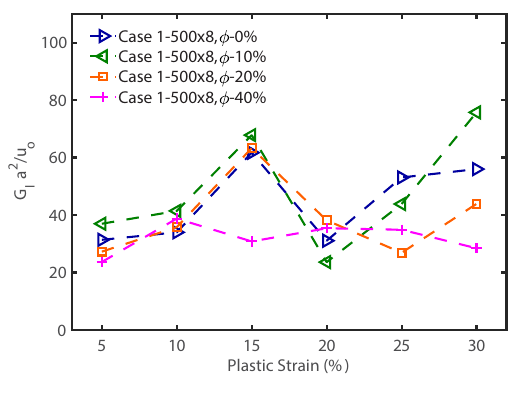}
    \caption{\textbf{Fracture energy of debonded samples of Case 1 (Set 2)}}
    \label{fig:Fracture}
\end{figure}

%\end{doublespace}
\bibliographystyle{unsrt}
\bibliography{bibliography.bib}
\end{document}